\documentstyle[aps,prb,preprint]{revtex}
\begin{document}

\title{Spin Glass and Antiferromagnet Critical Behaviour in a Diluted
  fcc Antiferromagnet}

\author{Carsten Wengel$^a$, Christopher L. Henley$^b$,
Annette Zippelius$^a$}

\address{$^a$Inst. f. Theoretische Physik,
Georg--August--Universit\"at, 37073 G\"ottingen, Germany}

\address{$^b$Department of Physics, Cornell University, Ithaca,
NY 14853}

\maketitle

\begin{abstract}
In this paper we report on a Monte Carlo study of a diluted
Ising antiferromagnet on a fcc lattice.
This is a typical model example of a highly frustrated antiferromagnet,
and we ask, whether sufficient random dilution of spins does
produce a spin glass phase. Our data strongly
indicate the existence of a spin glass transition for
spin--concentration $p<0.75$: We
find a divergent spin glass susceptibility and a divergent
spin glass correlation length, whereas the antiferromagnetic correlation
length saturates in this regime. Furthermore, we find a first order
phase transition to an antiferromagnet for $1\ge p>0.85$, which becomes
continuous in the range $0.85>p>0.75$.
Finite size scaling is employed to obtain critical exponents.
We compare our results with experimental systems as diluted
frustrated antiferromagnets as ${\rm Zn_{1-p}Mn_{p}Te}$.
\end{abstract}


\section{Introduction}

The family of diluted magnetic semiconductors (DMS) of the general form
${\rm A^{II}_{1-p}Mn_{p}B^{VI}}$ encompasses a wide variety of alloys
which have been under extensive investigation during the past 15
years. These alloys form a zincblende structure, where the
${\rm B^{VI}}$ element occupies one fcc lattice while
${\rm A^{II}}$ and Mn share the second fcc lattice.
One fundamental aspect of research has focused on the magnetic
order of these systems, since they offer practical examples of
strongly frustrated, randomly diluted three dimensional fcc Heisenberg
antiferromagnets (afm) with dominant nearest neighbour
interaction\cite{giebultowicz88,furdyna87}.

In this paper we present results of a Monte Carlo study of a diluted
frustrated Ising model on a fcc lattice given by the Hamiltonian
\begin{equation}
{\cal H}=-J\sum_{<i,j>}\epsilon_i\epsilon_j s_i s_j,
\ \  \epsilon_{i} = \left\{ \begin{array}{r@{\quad}l} 1 &
\hbox{with prob.}\quad p \\ 0 & \hbox{with prob.}\quad
1-p. \end{array} \right.
\end{equation}
Here, $J$ is the coupling constant, which we will set $J=-1$ henceforward,
and $p\in[0,1]$ is the probability
that a lattice site $i$ is occupied with an Ising--spin $s_i$.
We are interested in the static properties of this model for different
dilution regimes. Besides the pure ($p=1$) and the slightly diluted
case ($p\sim 1$), that has already been studied by MC
simulation\cite{grest79,phani80,polgreen84} and
other methods\cite{mackenzie81,styer85}, we concentrate our interest on
the strong dilution regime, which has only been investigated in
experimental Heisenberg--systems as mentioned above.
Although in our work we perform a simulation of an Ising system
we find that some typical DMS results can be reproduced with
our simplified model.

%%%\subsection{Experimental motivation}

Neutron diffraction experiments of thin ${\rm Zn_{1-p}Mn_{p}Te}$
films for $p\in[1.0,0.85]$)
revealed a first order phase transition to an antiferromagnetically
ordered state of ``Type--III''\cite{giebultowicz93}. At approximately
$p=0.85$ a tricritical point is encountered, where the phase transition
becomes continuous. The antiferromagnetic order in the regime
$p\in[0.75,0.85]$ is truly long range, however, below $p\approx0.75$
a transition to a short range ordered phase seems to appear
\cite{giebultowicz93,giebultowicz90}.

Below $p=0.7$ most experimental results
have led to the view, that one encounters a transition to a
spin--glass--like phase at a fairly well defined temperature $T_c$.
Characteristic spin glass features are: (i) remanence
effects in the frozen state\cite{mcalister84,geschwind88}, (ii) a
pronounced cusp in the susceptibility around $T_c$\cite{galazka80}
with strong frequency dependence of
the cusp temperature\cite{geschwind88}, (iii) absence of long range spin
order as observed by magnetic neutron
diffraction\cite{giebultowicz90}, (iv) dynamic scaling near $T_c$ of
frequency-dependent response
function\cite{zhou89,mauger90,geschwind90,leclercq93} and --- most
importantly --- (v) a divergent nonlinear
susceptibility around the cusp temperature\cite{mauger89}.

On the other hand, the antiferromagnetic
correlation length $\xi_{\mbox{\scriptsize AFM}}$ grows continuously
with decreasing temperature until it
saturates at the cusp temperature to an enormously large value as
high as 70 {\AA} at $p=0.7$; it is only below $p=0.4$ that short range
afm order disappears.
In the intermediate dilution range $p\in(0.4,0.7)$ the spin
glass interpretation has been questioned, and it was suggested
briefly\cite{geschwind88} (on the basis of an ``activated'' scaling
analysis) that the equilibrium transition at $T_c$ was to the
antiferromagnetically ordered state of a random--field system.

This motivates a numerical study of a diluted antiferromagnet,
in which we can observe the interplay of strong afm local order with
spin glass order, and can measure the quantities now considered
to be the signatures for spin-glass transitions.
Because Heisenberg model simulations demand more computer time,
and because of the more convoluted controversies regarding
the existence of a sharp phase transition for continuous spins,
we adopt here an Ising model; in addition, we have retained
only the first--neighbour exchange interaction.

%%%\subsection{Outline}

The paper is organized as follows:
Section~\ref{sec:theory} is concerned with the main theoretical arguments
that guide our expectations for the results of our simulations in the
distinct dilution regimes. In particular, we discuss the possible
universality class of the proposed spin glass phase.
Section~\ref{sec:tech} describes technical aspects of our simulations,
in particular the equilibration criterion.
In section~\ref{sec:first-order-afm}--\ref{sec:spin-glass-order}
of our report we shall present representative
data for the distinct dilution regimes. In
section~\ref{sec:first-order-afm}  we concentrate
on the pure case and on weak dilution ($p\sim 1$), where we
investigate how the order of the transition is being modified by
disorder in form of stochastically removing spins from the lattice.
Section~\ref{sec:cont-afm} is
concerned with the regime $p=0.8$, where a continuous phase transition
with afm ordering is found; critical exponents are determined by
finite size scaling. Section~\ref{sec:spin-glass-order} investigates
the intermediate and strong
dilution range ($p\in[0.3,0.7]$), where the question of the magnetic
ordering is our main concern.
A summary of the results, a comparison to experimental systems and
our final conclusions will be given in section~\ref{sec:conclusion}.

\section{Theoretical background}
\label{sec:theory}

In this section, we collect the qualitative expectations for all the
different concentration regimes expected in the phase diagram.
As with the results in the subsequent sections, we begin by reviewing the
pure case and proceed in the direction of greater dilution.
The global phase diagram is qualitatively similar to those conjectured
for {\it vector} spins in
Ref.~\onlinecite{henley89} and
Ref.~\onlinecite{giebultowicz86}, except of course that distinctive
collinear and noncollinear phases cannot exist in the Ising case.

\subsection{Pure fcc Ising antiferromagnet}

The pure Ising antiferromagnet on a fcc lattice has been extensively
studied both analytically and with simulations. Each spin has 12
nearest neighbours which in the ground state can only satisfy 8 bonds,
4 of them being always violated. This effect of frustration, which follows
from the triangles in the fcc lattice, leads to a large ground state
degeneracy\cite{danielian64} of the order ${\cal O}(2^L)$,
where $L$ denotes the linear extent of the system.
Thus the entropy per spin is zero as $L \to \infty$.

At small temperatures in this system,
thermal fluctuations generate free energy terms which have the
same effect as ferromagnetic
second neighbor interactions: this favors the ``Type--I''
afm order, meaning that the system orders into one of the
six periodic ground states with a $\langle 100\rangle$ type ordering
wavevector\cite{mackenzie81}.
This is an example of what Villain called ``ordering  due to disorder''
\cite{villain79}.

The discrete choice between the three $\langle 100\rangle$ type
directions suggests
a similarity in behavior to the 3-state Potts model, which has a
weakly first order phase transition in three dimensions.
Indeed, the $4-\epsilon$ renormalization group
predicts a first-order phase transition\cite{RG1storder}.
Simulations \cite{phani80,polgreen84,binder81} and series
expansions\cite{styer85} confirm that the phase transition is at finite
temperature and is of first order.

The antiferromagnetic state may be handled quantitatively by constructing
(in the spirit of Ref.~\onlinecite{bak76})
a three-component staggered-magnetization order parameter
$\mbox{\boldmath $m$}^{\mbox{\tiny\dag}} =
(m^{\mbox{\tiny\dag}}_1,
m^{\mbox{\tiny\dag}}_2,
m^{\mbox{\tiny\dag}}_3) $
with components
\begin{equation}
\label{staggmag}
m^{\mbox{\tiny\dag}}_\mu=\frac{1}{N}
\sum_j \exp({-{\rm i}\,\mbox{\small\boldmath $r$}_{\!j}\cdot
\mbox{\footnotesize\boldmath $k$}_\mu})\:s_j
\end{equation}
for $\mu=1,2,3$;
here the ordering wave vectors
are $\mbox{\footnotesize\boldmath $k$}_1 =(\pi,0,0)$ and cyclic
permutations (We have taken the lattice constant to be unity.).

\subsection{Weak dilution: random-field effects}

Dilution in frustrated systems (without any external field)
couples to the order parameter as a random field
does in a ferromagnet\cite{fernandez88,henley89}.
Take the case of rather weak dilution, which justifies
assuming (as a sort of variational state)
one of the six $\langle 100\rangle$ type ground states.
Consider the effect of strengthening one {\it bond} lying within the
$xy$ plane: it will favor the four states with
(100) and (010) ordering wavevectors and
disfavor the two with (001) wavevectors, since the bond in question is
violated in the (001) states.  The effect is much like a random field,
except that it does not destroy the global up/down symmetry.
In our case of {\it site} dilution, the random-field-like effects of removing
an isolated site cancel each other; however, removing a {\it pair of sites}
has the same effect as would strengthening the bond between
them\cite{fernandez83,fernandez88}.

Quite generally, when the random field is sufficiently
strong, the first-order transition is converted to a continuous
one\cite{berker91}.
In the present context, since the effective random field grows with dilution,
this argument predicts a tricritical point: the ordering transition is
first-order for $p > p_{tri}$ but
becomes continuous for $p < p_{tri}$\cite{pottsmulti}.

For $p< p_{tri}$ the transition from the paramagnet
is expected to be a novel universality class\cite{berker91}.
It would seem plausible if its dynamic scaling behavior
were of the ``activated'' type,
as in the random-field Ising model.
No experiments have tested this, however (The materials in
this concentration range, roughly $0.7 < p < 0.85$, can be grown
only as thin epitaxial slabs, meaning that very little signal
is available for susceptibility experiments.).
A claim was made that ``activated'' scaling could fit the data
\cite{geschwind88}
for lower values of $p$, which we would identify as the spin-glass phase,
but this was quickly corrected\cite{zhou89,geschwind90,leclercq93}.

When the ``effective random fields'' are sufficiently strong,
the antiferromagnetic order disappears  and is replaced by
spin-glass order at a critical value $p_*$\cite{twopstar}.
Note that this threshold to propagate afm order is far above the
$p_c$, the geometrical percolation threshold for propagating
connectivity of nearest-neighbor sites\cite{shnidman84}
(On the fcc lattice, $p_c= 0.195$\cite{essam71}.).

\subsection {Spin glass phase}

Any spin glass, by definition, requires random frustration.
This can be realized by dilution of a uniformly frustrated antiferromagnet,
as in the present case,
just as well as by a random mix of ferromagnetic and antiferromagnetic
interactions\cite{villain79}.
Indeed, the effective coupling between two spins may be ferromagnetic
or antiferromagnetic depending on how
intervening sites happen to be occupied.
Of course, this spin glass state is expected to show residual
short--range correlations as opposed to the cases of the $\pm J$ or Gaussian
random bond distributions, where symmetry implies that
$[\langle s_i s_j\rangle]=0$ if $i\neq j$.

It is intriguing that this ``spin glass'' state,
which occurs below $p_*$, from the viewpoint of the
antiferromagnetic phase,
is the same as the disordered domain state
which is favored by the ``effective random fields'' mentioned
above\cite{afmsgclass}.
This state is different from the familiar random--field disordered state
(and  similar to the usual spin glass state) because the
global up/down symmetry is preserved; consequently, for $p< p_*$
there is still a true phase transition in which this symmetry is
broken\cite{percorder}.

\subsection{Universality}

In a concentration range $(p_c', p_*)$, the ground state is
presumed to be a spin glass.
Spin glass investigations tend still to be preoccupied with the issue
of the existence of a transition as a function of dimension, external field,
and spin type.
Indeed, it is still unsettled whether the $d=3$ Ising spin glass really has a
transition at finite temperature, or whether it is at the lower critical
dimensionality\cite{bhatt88,marinetti94}.
Rather little has been done to test the universality of the
critical exponents, as almost all simulations have used simple cubic
(sc) lattices with the discrete $\pm J$ distribution of random bonds.
Monte Carlo and series studies for the
fcc lattice with $\pm J$ bonds \cite{hetzel93}
gave values of the spin-glass exponents $\eta$, $\nu$, and $\gamma$
consistent with the sc $\pm J$ model;
so did {\it diluted} $\pm J$ model on a simple cubic lattice\cite{shapira94}
and (modulo large error bars)
Gaussian-distributed random bonds on the simple cubic lattice
\cite{bhatt88}.
The above results are consistent with universality. However, it has
also been proposed that $\eta$ is more negative and the Binder cumulant is
larger for Gaussian bond randomness than for $\pm J$
randomness\cite{bernardi94}; presumably the diluted fcc is more
similar to the latter model, since its discrete randomness
generically allows exact degeneracy of ground states.

\subsection{Theory of $p_c'$ (spin glass near percolation)}

We now consider the approach $p \to p_c'$,
where the spin glass long--range order finally disappears.
In this regime, the order is just barely propagating along tortuous,
effectively one--dimensional paths, and consequently we
expect $T_c \to 0$ (exponentially) as
$p \to p_c'$\cite{bray87,shapira94}.

Note that $p_c' > p_c$.
In frustrated models with discrete bond distributions,
such as the present case,
two portions of a connected cluster might be
connected by (say) two chains of bonds,  each canceling  the other
and allowing one portion to be flipped relative to the other portion
at no cost in energy; for propagation of order, it is as if no chain
existed, i.~e., the effective concentration of bonds is lowered by
the cancellations.

\section{Technicalities}
\label{sec:tech}

We use the single spin flip Monte Carlo Metropolis algorithm in our
simulations. Spins are updated sequentially and randomly.
Periodic boundary conditions are imposed, limiting the
possible lattice sizes to even numbers. Spins are represented on a
cubic lattice with next nearest neighbour interactions to obtain a fcc
lattice. Therefore, every lattice of linear size $L$ contains
$N=L^3/2$ sites. We simulated lattice sizes $L=4,6,8,10$ with
$M\approx 120$ realizations of the disorder and $L=16$ with $M=40$.
We investigate the model in the concentration range $p\in[0.3,1.0]$.

A standard criterion by Bhatt and Young\cite{bhatt85} was
applied to test the equilibration of the systems throughout the
whole simulation, where we observe a continuous phase transition.
The procedure is to obtain two estimates of the spin glass phase
indicator, the spin glass susceptibility
\begin{equation} \label{chi-sg}
\chi_{\mbox{\scriptsize SG}}
=\frac{1}{N}\sum_{ij}\left[\langle s_is_j\rangle^2\right]
\end{equation}
Here and later, the brackets $\langle\cdots\rangle$ denote
thermal averaging and $\left[\cdots\right]$ configurational averaging.
We obtain $\chi_{\mbox{\scriptsize SG}}$ by calculating the second
moment of the spin glass order parameter defined in the two
alternative ways, (i) as the overlap
\begin{equation} \label{overlap}
q_{12}(t,t_0)=\frac{1}{N}\sum_i s_i^{(1)}(t_0+t)\, s_i^{(2)}(t_0+t)
\end{equation}
and (ii) as the autocorrelation (self--overlap)
\begin{equation}
\label{autocor}
q_{t^{\prime}}(t,t_0)=\frac{1}{N}\sum_i s_i(t_0+t)\,
s_i(t_0+t+t^{\prime}).
\end{equation}
Here, $s_i^{(1)}$ and $s_i^{(2)}$ denote two sets of spins (replicas)
with the same set of occupied sites and uncorrelated initialization
and $t_0$ is the time initially used for equilibration.

With these definitions, we can compute two estimates of
$\chi_{\mbox{\scriptsize SG}}$ as follows, i.~e.,
\begin{equation} \label{chi-sg-overlap}
\chi_{\hbox{\scriptsize SG}}^{(a)}=\left[\left\langle\frac{1}{N}
\left(\sum_i s_i^{(1)}(t_0+t)\,s_i^{(2)}(t_0+t)\right)^{\!\!2}
\right\rangle_{\!\!\tau}\right],
\end{equation}
respectively the four--spin correlation function,
\begin{equation}
\chi_{\hbox{\scriptsize SG}}^{(b)}=
\left[\left\langle\frac{1}{N}\left(\sum_i
s_i(t_0+t)\,s_i(2t_0+t)\right)^{\!\!2}
\right\rangle_{\!\!\tau}\right],
\end{equation}
where
$\langle\cdots\rangle_{\tau}=\frac{1}{\tau}\sum_{1}^{\tau}
(\cdots)$, and $\tau=t_0$.

The equilibration time $t_0$ was raised on a logarithmic scale and we
only accepted a run, if both estimates of
$\chi_{\mbox{\scriptsize SG}}$ agreed after this time within
certain limits, typically of the order of 5\% of their joint mean
value. The longest runs
performed were up to $2*10^6$ Monte Carlo steps per spin (MCS).
Most of the simulations were performed on HP workstations at the
Institut f\"ur Theoretische Physik, G\"ottingen and on the
Intel--Paragon parallel computer at the H\"ochstleistungsrechenzentrum
J\"ulich. The program was parallelised by using {\it PVM 3.2} software
\cite{pvm93} in order to simulate many systems simultaneously.

\section{First order antiferromagnetic transition (weak dilution)}
\label{sec:first-order-afm}

In this section we investigate the pure afm and the slightly diluted
regime, i.~e., $p\in[0.85,1.0]$. We wish to determine the order of the
phase transition and how the order of the transition is changed by
introducing disorder into the system in form of slight stochastic dilution.
The magnetic order in this regime is clearly antiferromagnet,
consistent with earlier simulations, and shall be closer examined
in section \ref{sec:cont-afm}.

The pure antiferromagnet on a fcc lattice is known to undergo
a temperature driven first order phase
transition\cite{grest79,phani80,binder81,polgreen84,styer85}, as
mentioned in the previous section.
Early Monte Carlo simulations by Grest and Gabl\cite{grest79} as well
as Giebultowicz\cite{giebultowicz86} reported a
change from a first order to a continuous phase transition upon
dilution.
Grest and Gabl located the tricritical point at a critical concentration
$p_{\hbox{\scriptsize tri}}=0.93$ using
Ising--spins, whereas Giebultowicz found a slightly lower
$p_{\hbox{\scriptsize tri}}=0.85$ with a Heisenberg--spin
simulation. However, in both of those simulations, no averaging over
the disorder was performed, so that we re--investigated this regime.

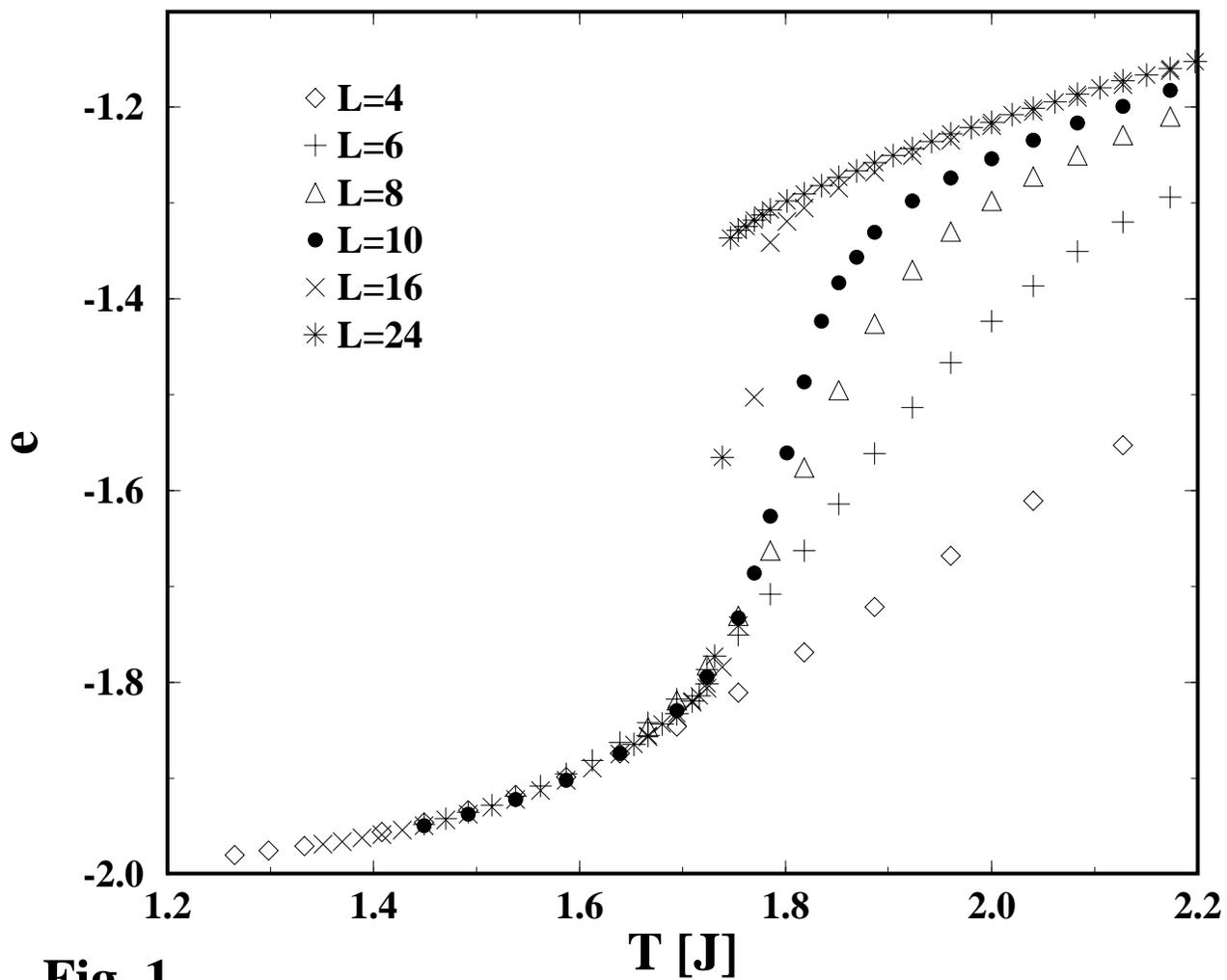
\begin{figure}
\caption{Internal energy density $e$ versus temperature for $p=1$.}
\label{energy-heat}
\end{figure}

An important quantity for a first order phase transition is the
internal energy density
\begin{equation}
[\langle e\rangle]=\left[\left\langle\frac{1}{2N}\sum_{<i,j>}\epsilon_i
\epsilon_j s_i s_j\right\rangle\right].
\end{equation}
At the critical temperature, this quantity indicates a first order phase
transition by a discontinuity (latent heat), which can be seen in
Fig. \ref{energy-heat}, where $[\langle e\rangle ]$ is plotted
versus temperature $T$; with increasing lattice size a pronounced
discontinuity at the transition temperature can be observed, revealing
clearly a first order transition.

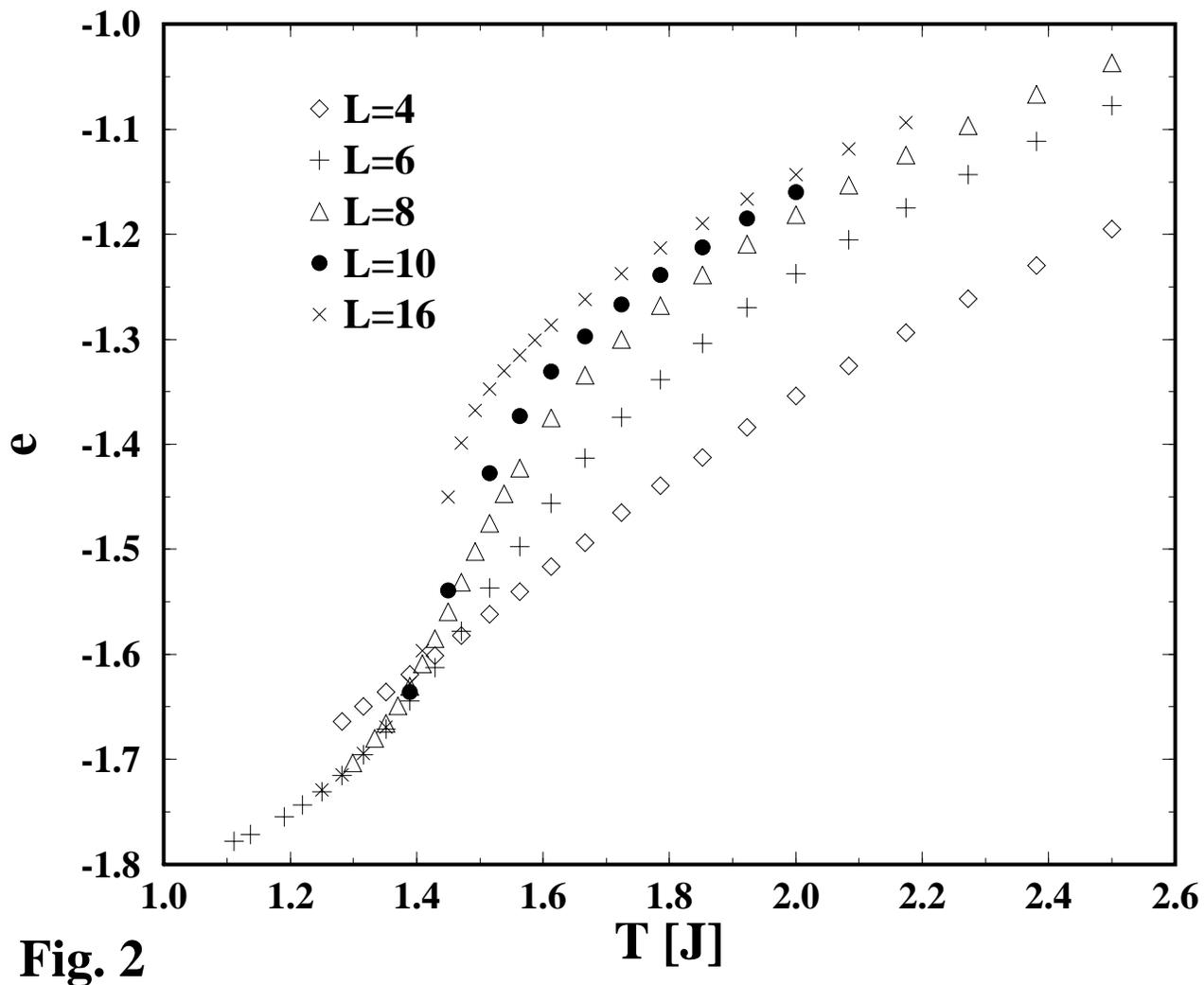
\begin{figure}
\caption{Internal energy density versus temperature for $p=0.9$.}
\label{energy-p=0.9}
\end{figure}

In Fig. \ref{energy-p=0.9} we present our data of the
internal energy density for $p=0.9$.
For this concentration the transition appears to be continuous. Apparently,
a tricritical point is hard to locate with these Monte Carlo
data, since it is not clear, whether in the limit of smaller
temperature steps and larger systems the transition will turn out to
be continuous or not. In order to determine the
tricritical concentration more efficiently we used a method first
introduced by Lee
\& Kosterlitz \cite{lee90}, which is based on a histogram algorithm
by Ferrenberg \& Swendsen\cite{ferrenberg88}.

\subsection{Histogram algorithm and finite-size scaling analysis method}

This method\cite{ferrenberg88} uses one long Monte Carlo
run to estimate the
free energy at several temperatures close to $T_c$. During the runs a
histogram of the internal energy density $e$ is accumulated .
This histogram serves as an
estimate of the equilibrium energy distribution (after correct
normalization)
\begin{equation} \label{ener-dist}
P_{\beta}(e)=\frac{1}{Z_{\beta}}\exp{(S(e)-\beta e)}.
\end{equation}
Here, $Z_{\beta}$ is the partition function at the inverse temperature
$\beta=1/k_BT$ and $S(e)$ is the entropy. Notice, that $P_{\beta}(e)$
(\ref{ener-dist}) is proportional to $\exp(-\beta F(e))$, where $F$
denotes the free energy. The distribution
$P_{\beta}(e)$ can now be used to generate the distribution (and
consequently $F(e)$) at a different inverse temperature
$\beta^{\prime}$ in the vicinity of $\beta$.

We are interested in the situation in which $F(e)$
has two minima $e_1$ and $e_2$
(i.e. (\ref{ener-dist})  has maxima at these energies).
The temperature at which $F(e_1,L)=F(e_2,L)$
is taken to be the effective critical temperature for the given size $L$.
Then we evaluate the ``gap'' $\Delta F=F(e_m)-F(e_{1,2})$
between the free energy  values at those two energies
and at the maximum $e_m$ in between them.
{}From a scaling analysis of $\Delta F$, one can identify whether a
phase transition is first order\cite{lee90}.

A state with $e\in [e_1,e_2]$ consists of a domain of
ordered and a domain of disordered phase coexisting,
separated by a $(d-1)$-dimensional interface
surrounding the droplet of minority phase.
Therefore, one can expand the free energy
\begin{equation}  \label{free-ener-exp}
F(e,L)=L^d f_0(e) + L^{d-1} f_1(e) + {\cal O}(L^{d-2}),
\end{equation}
where the bulk free energy density $f_0$ is minimal and constant
for $e\in [e_1,e_2]$ and the surface term $f_1$ is maximal at
$e_1<e_m<e_2$.
Expansion (\ref{free-ener-exp}) is valid for any first order phase
transition, as long as the correlation length $\xi<L$. At the critical
temperature $\xi$ remains finite, so that for sufficiently large
$L$ the appearance of a free energy gap
$\Delta F$ indicates a first order
transition; for small $L<\xi$ the free energy is dominated by the bulk
term. As the system approaches
a tricritical point, $\xi$ grows and the double minima structure can
only be seen for large $L$.
At a tricritical point and beyond it, the phase transition is continuous
and hence there is no double minimum structure for any $L$.

\subsection{Results}

To obtain good statistics, we took histogram data
every $10^{th}$ MCS for $6\times 10^6$ MCS,
averaging over 16 realizations of the disorder.

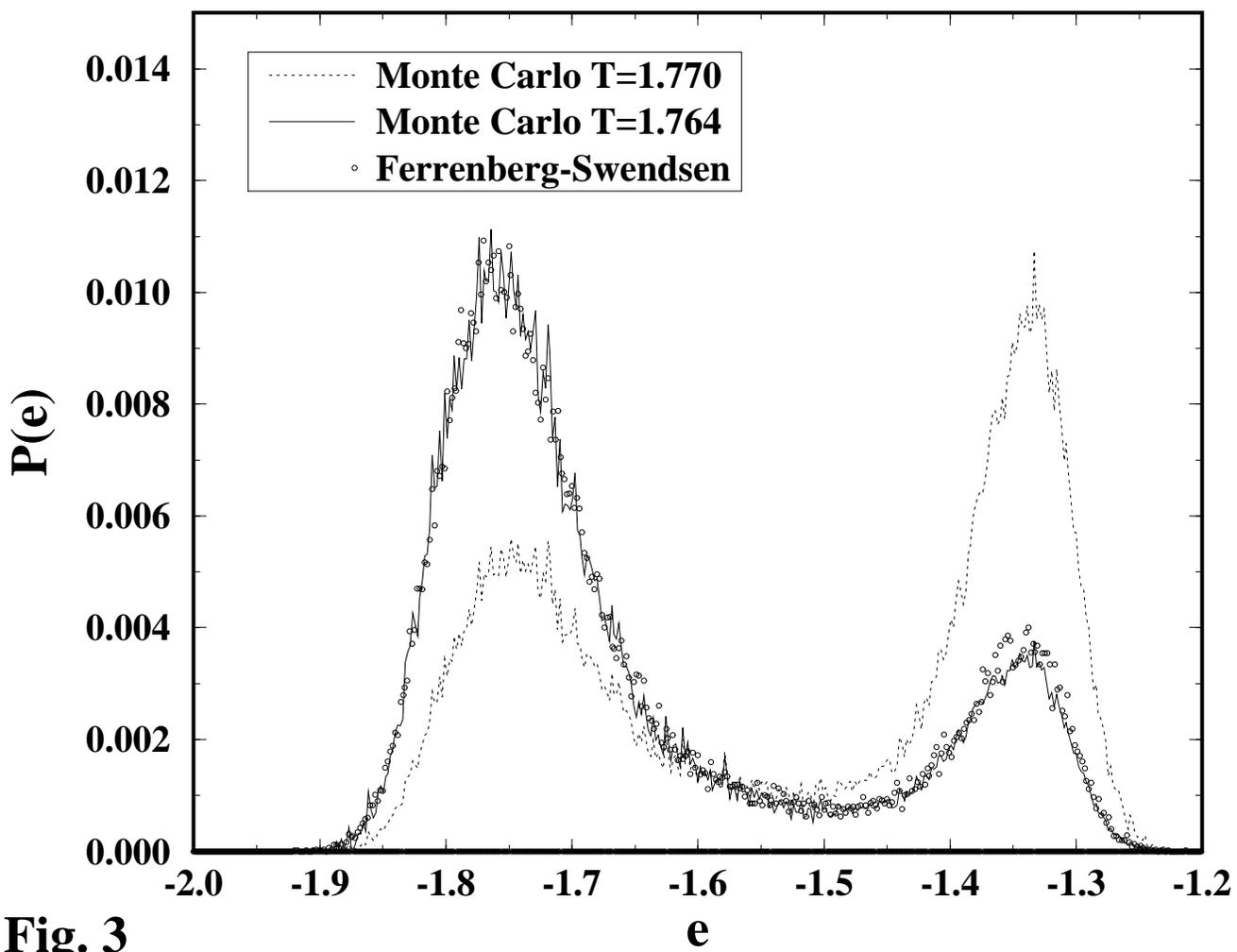
\begin{figure}
\caption{The lines show two energy density distributions $P(e)$ versus
        energy density $e$ for MC--simulations at
        two different temperatures, $p=1.0$ and $L=16$. The left
        peak corresponds to an antiferromagnetically ordered phase
        (low energy), the right peak to a paramagnetic phase, divided
        by an energy gap (latent heat). The points are calculated
        according to the Ferrenberg--Swendsen
        method from the high temperature distribution and agree well
        with the full line from low temperature MC--simulation.}
\label{hist-2temp}
\end{figure}

Fig.~\ref{hist-2temp} shows two histograms of
the internal energy density for $p=1$ after
normalization for two different temperatures (full and dashed line).
The dots were produced by transforming the $T=1.77$ data set (dashed line)
to the lower temperature $T=1.74$. The distribution
depends sensitively on the chosen temperature; the Monte Carlo data
at $T=1.74$ and the transformed data from $T=1.77$ show good
agreement, indicating, that the equilibrium distribution has been well
estimated. The simulation temperature in Fig. \ref{hist-2temp} was chosen
very close to the critical temperature. Therefore, as noted above,
the distribution clearly has two maxima at $e_1$ and $e_2$, which are
equivalently minima of $F(e)$.

If we define the effective critical temperature $T_c(L)$ by the
equality of the two maxima $P_{\beta}(e_1)$ and $P_{\beta}(e_2)$ and
extrapolate $T_c(L\to\infty)$, we obtain $T_c(p=1.0)=1.716(5)$. This
is slightly lower than the values from Ref.~\onlinecite{polgreen84}
($T_c=1.736(1)$) and Ref.~\onlinecite{styer85} ($T_c=1.746(5)$). This
discrepancy may be due to using different definitions of
``finite--size $T_c$'' in the respective references.

While for $p=1$ the energy gap appears already at $L=10$, for stronger
dilution it can only be seen at larger system sizes,
indicating the growing correlation length as one approaches the
tricritical point. If a gap appeared at one system size, it continued
to be present and in fact grew for larger sizes.
The scaling behaviour according
to eq. (\ref{free-ener-exp}), i.~e., $\Delta F(L) \propto L^2$,
is fulfilled within the error margins. In the diluted case
we found considerable fluctuations of the
energy gap size depending on the realization. At $p\sim 0.9$ for our
largest systems ($L=22$) certain realizations were found, that showed
none of the typical two peak structure, while others still showed a
small gap.

In order to determine the tricritical dilution
$p_{\mbox{\scriptsize tri}}$, we evaluated the percentage of
realizations with gap ($=Y$) with decreasing
concentration of spins. For our largest systems ($L=22$),
at $p=0.91$ $Y=90\%$ of the realizations still showed a double peak,
at $p=0.9$ only 40\%, at
$p=0.89$ 10\% and for $p=0.88$ no double peaks were found at all,
i.~e., all systems showed Gaussian peaks at $T_c$. From extrapolating
$(T_c(L),p_{\mbox{\scriptsize tri}}(L))$ determined by the condition
$Y=0.5$ for different system sizes (see dots in
Fig. \ref{phase-diag}), we estimated
$p_{\rm tri}=0.85\pm 0.03$ in accordance with
Ref.~\onlinecite{giebultowicz86}.
To determine the triciritcal concentration more exactly would require
a precise scaling analysis based on more extensive data,
which was beyond the scope of this work.

\section{Continuous antiferromagnetic transition ($p=0.8$)}
\label{sec:cont-afm}

In this section we concentrate on simulations performed for $p=0.8$.
Our aim is to check, whether at $T_c$ the system orders into an
antiferromagnetic state and, since we are below the tricritical
concentration we measure critical exponents via finite size scaling
close to the continuous phase transition, that we encounter.

\subsection{Quantities analysed}

For the antiferromagnetic phase, the staggered magnetization
``vector'' $\mbox{\boldmath $m$}^{\mbox{\tiny\dag}}$
(see eq.~\ref{staggmag}) is the appropriate order parameter.
Thus, we calculate the second moment
\begin{equation} \label{stag-mag}
m^2=[\langle\mbox{\boldmath $m$}^{\mbox{\tiny\dag}}\cdot
\mbox{\boldmath $m$}^{\mbox{\tiny\dag}}\rangle].
\end{equation}
For $T>T_c$ this is proportional to the staggered susceptibility
$\chi^{\mbox{\tiny\dag}}=N m^2$,
which we analysed by using the finite size scaling form
\begin{equation} \label{chi-dagg-scale}
\chi^{\mbox{\tiny\dag}}(L,T)=L^{2-\eta}\,\tilde{\chi}^{\mbox{\tiny\dag}}
(L^{1/\nu}(T-T_c))
\end{equation}
to extract the critical exponents $\eta$ and $\nu$ and the critical
temperature $T_c$.

Since we are mainly interested in the magnetic order of the different
dilution regimes, we also calculated correlation functions.
To save computer time, we Fourier transform
the lattices to $k$--space and calculate the Fourier transformed
correlation function, i.~e.,
\begin{equation}
G(\mbox{\small\boldmath $k$})=[\langle |
s_{\mbox{\footnotesize\boldmath $k$}} |^2 \rangle ]=
\frac{1}{N}\sum_{ij} \exp (-{\rm i}\:
\mbox{\small\boldmath $r$}_{\!ij}
\cdot\mbox{\footnotesize\boldmath $k$})\: [\langle s_i s_j\rangle ].
\end{equation}
We applied the Fast Fourier algorithm to the (most relevant)
$L=16$ systems only and calculated the
correlation function along the three $\langle 1 0 0\rangle$ directions.
For an antiferromagnet, the three $k=\pi$ modes of $G$ should
yield the static staggered
susceptibility, which we used as a consistency check.
The correlation length $\xi$ can be
extracted from the knowledge of the scaling form of $G$, i.~e.,
\begin{equation}
G(k)=\frac{1}{k^{2-\eta}}\tilde{G}(k\xi).
\end{equation}
The scaling factor $\tilde{G}$ has the following asymptotics: For
$k>0$ and $T\to T_{c}^+$, $\tilde{G}(k\xi)\to1$ and
and for $T>T_c$ and $k\to0$ (in the afm case $k=\pi-k^{\prime}$ with
the limit $k^{\prime}\to\pi$)
\begin{eqnarray} \label{cor-scal-susc}
G(k\to0) & \sim & \frac{1}{k^{2-\eta}}(k\xi)^{2-\eta}=\xi^{2-\eta}\\
\nonumber
& \sim & (T-T_c)^{-\nu(2-\eta)}=(T-T_c)^{-\gamma}\\ \nonumber
&\sim&\chi.
\end{eqnarray}
In eq. (\ref{cor-scal-susc}) we have used the scaling form of the
correlation length $\xi\sim(T-T_c)^{-\nu}$ and the scaling law
$\gamma=\nu(2-\eta)$. With this knowledge of the asymptotic form of
$G$ we chose the Ansatz
\begin{equation}  \label{cor-fun-fit}
G(k)=\frac{1}{k^{2-\eta}}\tilde{G}(k\xi)=\frac{A}{k^{2-\eta}+
\xi^{\eta-2}},
\end{equation}
that we used to fit our data of the correlation function in
order to obtain an estimate for $\xi$ and $\eta$. The constant $A$ has
been introduced as the amplitude of the correlation function.

Finally we shall analyse the heat capacity
\begin{equation}
C=\frac{N}{T^2}\left[\langle e^2\rangle - \langle e\rangle^2 \right].
\end{equation}
This quantity indicates a continuous
phase transition to an ordered state by a weak divergence at the
critical temperature. We use the finite size scaling form
\begin{equation} \label{heat-scale}
C(L,T)=L^{\alpha/\nu}\tilde{C}(L^{1/\nu}(T-T_c))
\end{equation}
to extract critical exponents $\alpha$ and $\nu$ as well as $T_c$.

\subsection{Results}

Our data of the staggered magnetization show an increasing
$m^2$ (eq.~\ref{stag-mag}) as $T$
approaches the critical temperature, becoming more pronounced for
larger lattice sizes. This contribution to the afm order parameter can
as well be seen in the divergent behaviour of the staggered
susceptibility (Fig.~\ref{stag-sus}).
The scaling analysis\cite{errorbars} of the staggered susceptibility yields
$T_c(p=0.8)=1.07\:(0.05)$, $\nu=0.51\:(0.1)$ and $\eta=0.05\:(0.1)$.
The data scale well over a wide temperature range with the exception
of the $L=4$ data and the errors in the exponents are within
acceptable limits.

The antiferromagnetic correlation length
$\xi_{\mbox{\scriptsize AFM}}$ is found to
grow continuously with decreasing temperature and reaches half
the lattice size at $T\approx1.2$, before the critical
temperature $T_c=1.07$. Since finite size effects become appreciable
at distances close to half the lattice size, we only included data
above $T=1.2$ for a fit of the scaling form $\xi\sim(T-T_c)^{-\nu}$.
We extracted $\nu=0.55$ by regression, which is in good
agreement with our previous result.
{}From fitting the correlation function to eq. (\ref{cor-fun-fit}), we were
also able to extract a second estimate of $\eta$; here, we found
$\eta=-0.04$ with a slow drift to $\eta\to-0.02$ for $T\to1.2$.

We do not perform a complete finite size scaling analysis of the
correlation function in this work, because it would require a substantial
amount of additional data for larger lattice sizes.
Therefore, our estimates for the exponents obtained from $G(k)$ are less
reliable than those obtained from finite size scaling.
Nevertheless, this analysis serves as an additional consistency check and
the critical exponents lie well within the error margins of those
exponents obtained via finite size scaling of the susceptibility.

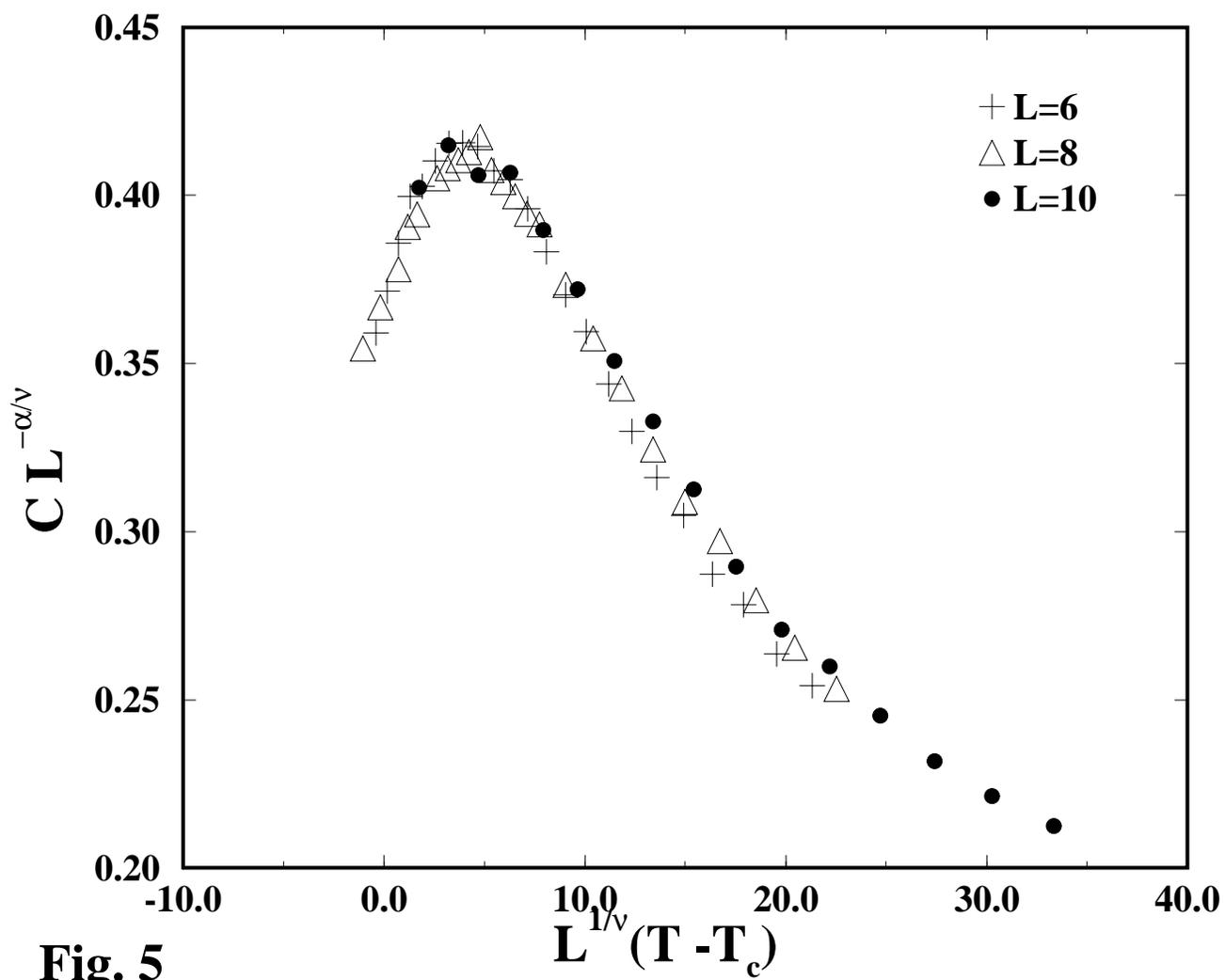
\begin{figure}
\caption{Scaled heat capacity $CL^{-\alpha/\nu}$ versus
         $L^{1/\nu}(T-T_c)$, $p=0.8$. Best scaling was achieved with
         $T_c=1.08\:(0.05)$, $\nu=0.57\:(0.1)$ and
         $\alpha=0.38\:(0.1)$.}
\label{heatp=8-scale}
\end{figure}

The heat capacity shows a weak divergence at the critical temperature
at $p=0.8$.
Fig. \ref{heatp=8-scale} displays the
best result of the scaling procedure according to
eq. (\ref{heat-scale}) with $T_c=1.08\:(0.05)$,
$\nu=0.57\:(0.1)$ and
$\alpha=0.38\:(0.15)$.
These exponent values satisfy the hyperscaling relation
\begin{equation}
d\nu=2-\alpha
\end{equation}
with $\nu=0.54$ in 3 dimensions.

In summary, the dilution regime $p=0.8$ exhibits
a continuous phase transition to an antiferromagnetically ordered state.
We obtained critical exponents using finite size scaling;
however, since $p=0.8$ is very close to the tricritical point, it
is quite possible that these are only effective exponents from
the crossover between the
mean field exponents of the tricritical
point ($\eta=0$, $\nu=1/2$ and $\alpha=1/2$) to whatever universality class
is appropriate for the continuous ordering transition.
With the data at hand this question has to remain open.

\section{Spin glass order}
\label{sec:spin-glass-order}

Upon further dilution we encounter a dramatic change in the magnetic
order of the system at the phase transition. We investigate the
question, whether below $p=0.8$ the system really orders into a spin
glass or if antiferromagnetic ordering can still be
found. First, we introduce the spin glass order parameter and
the spin glass susceptibility.
Then we discuss our data at $p=0.7$, and proceed with results from
simulations of stronger diluted systems.

\subsection{Theory and quantities measured}

Besides the quantities that have been introduced in the previous
section, we also measured the standard indicator of a spin glass
transition, the spin glass susceptibility
$\chi_{\mbox{\scriptsize SG}}$ (see eq.~(\ref{chi-sg})).
For an infinite system a spin glass transition is signalled by a
divergence of $\chi_{\mbox{\scriptsize SG}}$ as $(T-T_c)^{-\gamma}$,
with $\gamma=(2-\eta)\nu$.
For our scaling analysis, $\chi_{\mbox{\scriptsize SG}}$ is computed as
the second moment of the overlap, defined by eq.~(\ref{overlap}) and
eq.~(\ref{chi-sg-overlap}).
We analysed our estimate of the spin glass susceptibility by using
it's finite size scaling form
\begin{equation} \label{scale-chi-sg}
\chi_{\mbox{\scriptsize SG}}(L,T)=L^{2-\eta}\,
\tilde{\chi}_{\mbox{\scriptsize SG}}(L^{1/\nu}(T-T_c)).
\end{equation}

Another important quantity, that is well--known in
the analysis of spin glass simulation data\cite{bhatt85}, is the Binder
cumulant\cite{binder79} of the spin glass order parameter
\begin{equation}  \label{binder-cum}
g=\frac{1}{2}\left(3-\frac{\left[\langle q^4\rangle\right]}{\left[
\langle q^2\rangle\right]^2}\right).
\end{equation}
It has the pleasant finite size scaling form
\begin{equation}
g(L,T)=\tilde{g}(L^{1/\nu}(T-T_c))
\end{equation}
with no power of $L$ multiplying $\tilde{g}$, which makes it very
valuable for precise scaling analysis.
The Binder cumulant (\ref{binder-cum}) is defined so that $0\le g\le 1$,
and above $T_c$, $g(L,T)\to0$ for $L\to\infty$. In particular,
the intersection of all $g(L,T)$ curves at some point provides an
accurate estimate of $T_c$.

To investigate a change of magnetic order further, we analysed the
correlation function again, this time also the spin glass
correlation function
\begin{equation}
G_{\mbox{\scriptsize SG}}(\mbox{\small\boldmath $k$})=
[\langle |q_{\mbox{\footnotesize\boldmath $k$}} |^2 \rangle ]=
\frac{1}{N}\sum_{i,j}\exp(-{\rm i}\,
\mbox{\small\boldmath $r$}_{\!ij}\cdot
\mbox{\footnotesize\boldmath $k$})\:
[\langle q_i q_j\rangle ].
\end{equation}
Here, we defined $q_i=s_i^{(1)}(t+t_0)s_i^{(2)}(t+t_0)$. The same fitting
procedure of our data was employed as in the previous section in
order to compute the respective correlation length, $\nu$ and $\eta$.

\subsection{Results for intermediate dilution ($p=0.7$)}

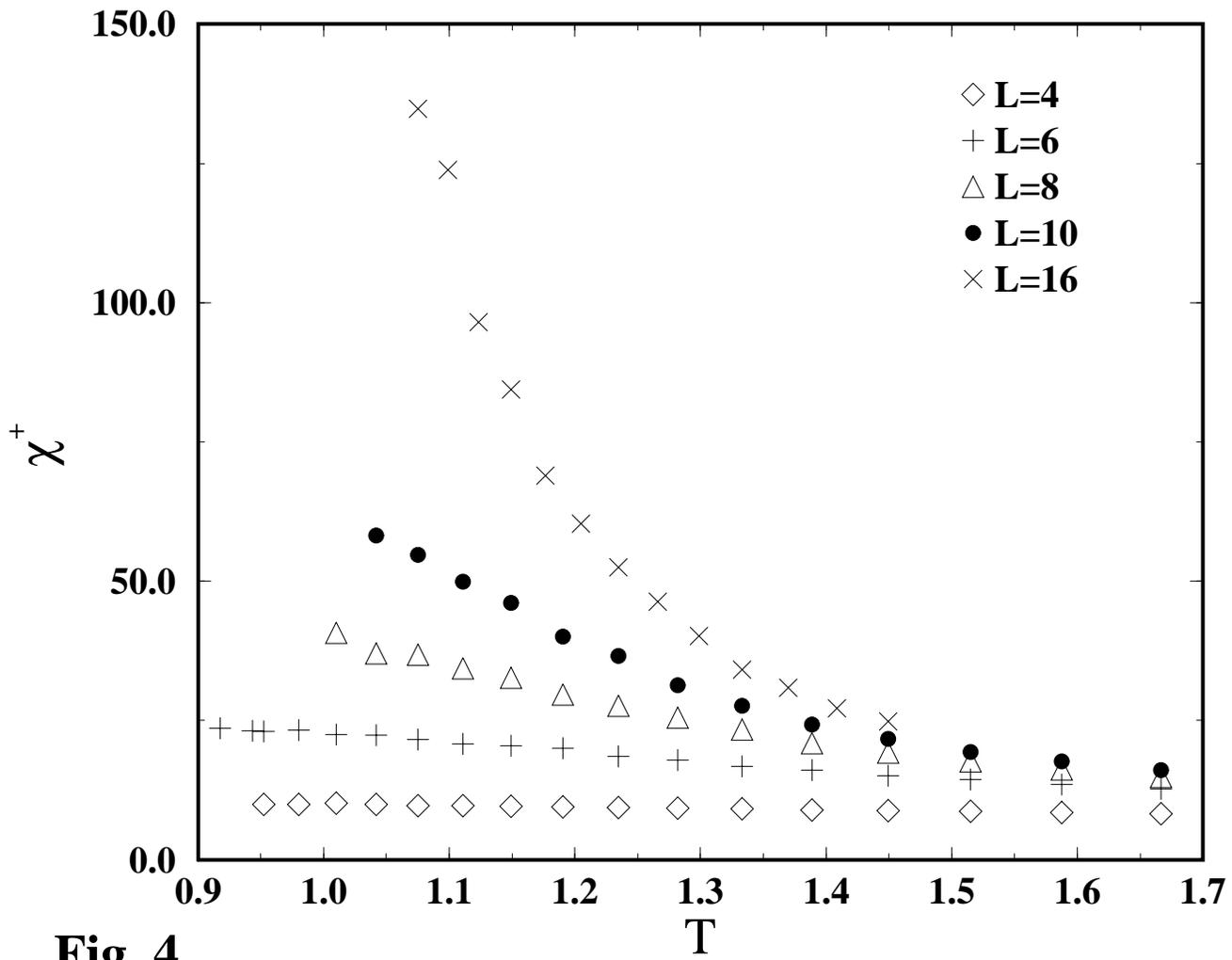
\begin{figure}
\caption{Staggered susceptibility $\chi^\dagger$ versus
        temperature, $p=0.8$.}
\label{stag-sus}
\end{figure}

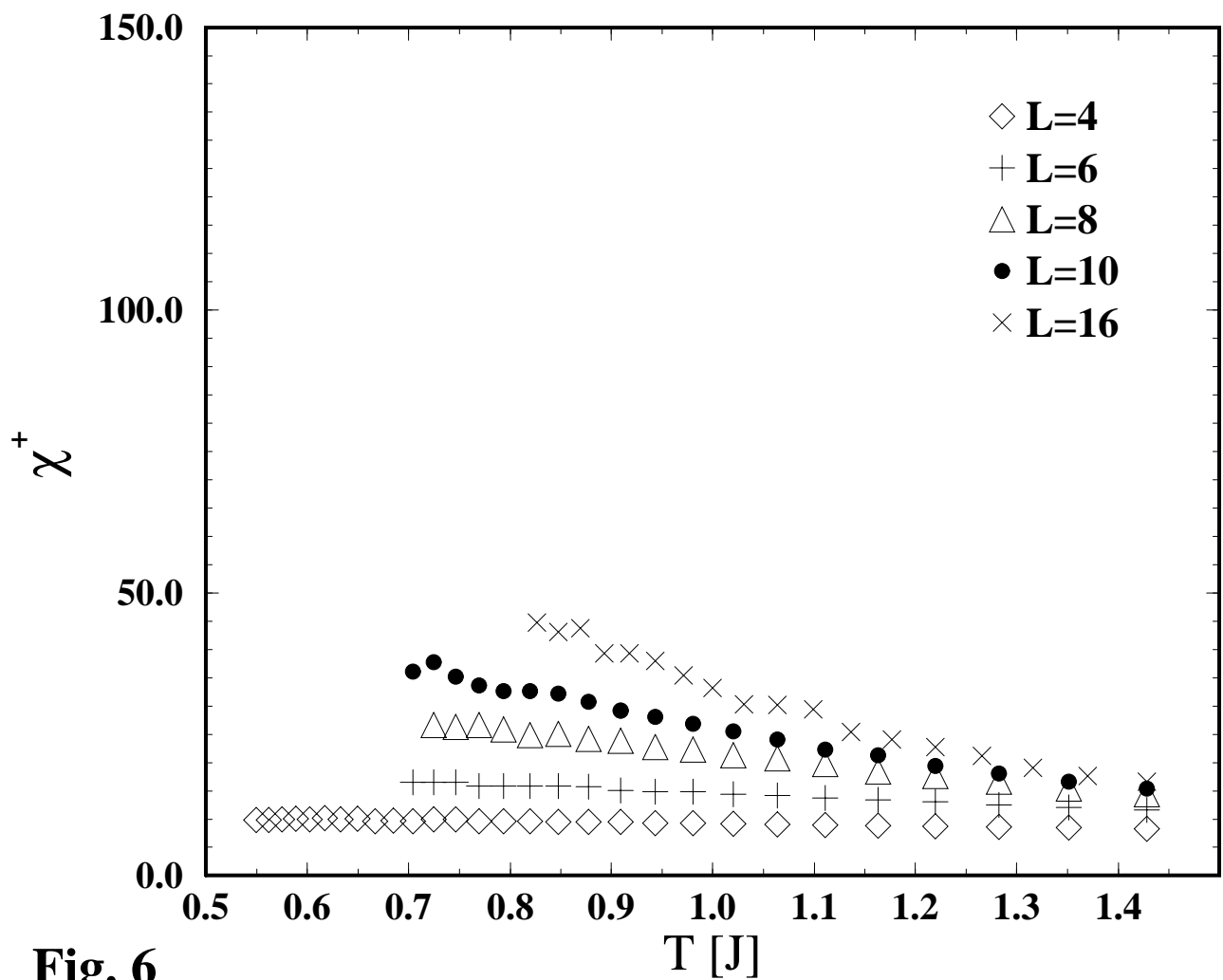
\begin{figure}
\caption{$\chi^\dagger$ versus temperature, $p=0.7$.}
\label{stag-susc=7}
\end{figure}

The (antiferromagnetic) staggered susceptibility
as shown in Fig. \ref{stag-susc=7}, is drastically reduced in
comparison to Fig. \ref{stag-sus}, and shows only
a small tendency to increase as $T$ decreases. Furthermore, scaling
according to eq. (\ref{chi-dagg-scale}) could not be achieved for
reasonable parameters.

On the other hand our data reveal a divergence of
$\chi_{\mbox{\scriptsize SG}}$ at $p=0.7$ of the same order of magnitude as
$\chi^{\mbox{\tiny\dag}}$ for $p=0.8$ in Fig. \ref{stag-sus}.
This divergence becomes particularly strong for larger lattice sizes
in the vicinity of $T=0.85$.
If $\chi_{\mbox{\scriptsize SG}}$ is the critical quantity for this
system then it should also scale according to eq. \ref{scale-chi-sg}.

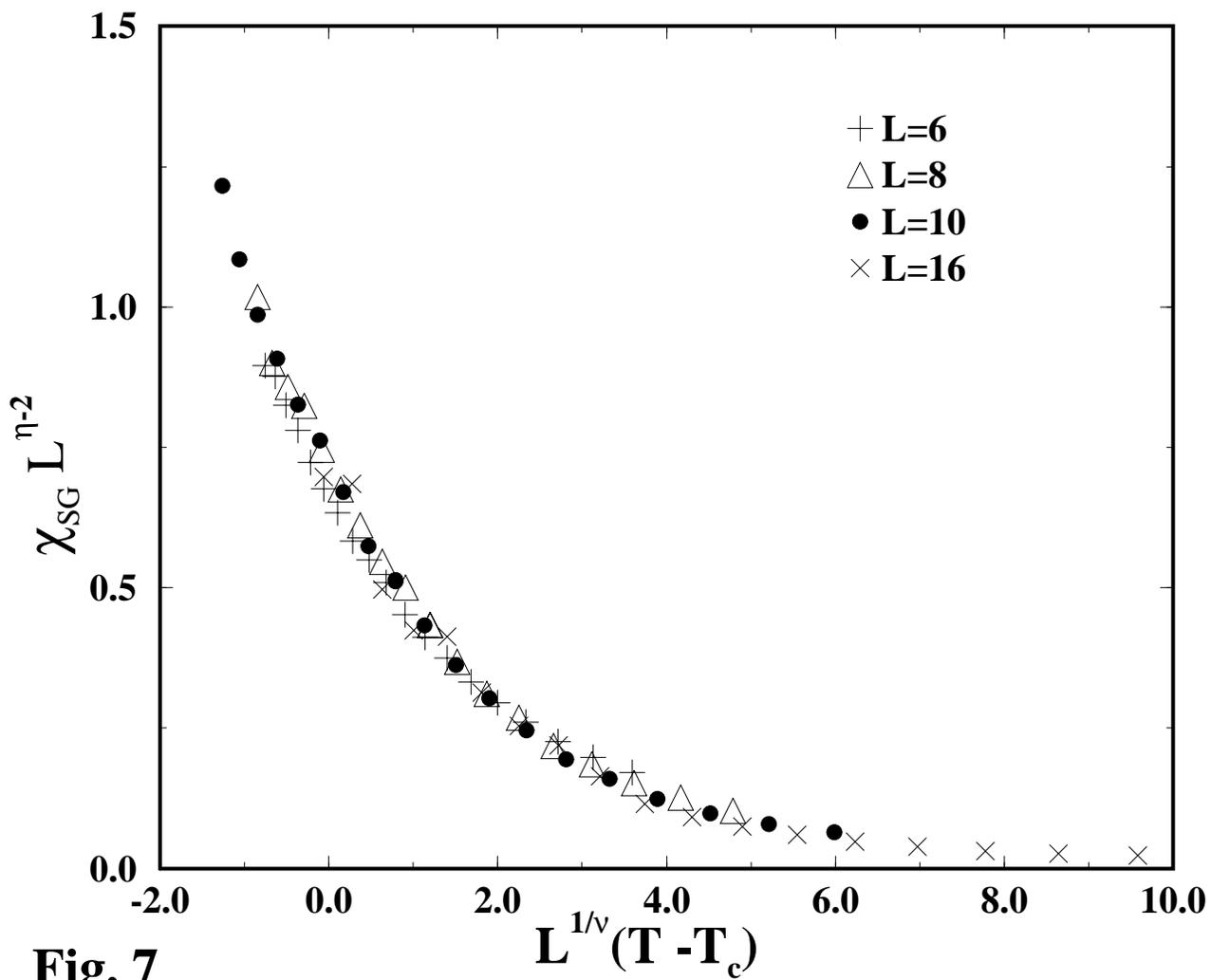
\begin{figure}
\caption{$\chi_{SG} L^{2-\eta}$ versus
         $L^{1/\nu}(T-T_c)$, $p=0.7$. Best scaling was achieved with
         $T_c=0.83 (0.05)$, $\nu=1.0\:(0.2)$ and $\eta=0.1\:(0.2)$.}
\label{chi-sgp=7-scale}
\end{figure}

A finite size scaling analysis of the spin glass susceptibility is
given in Fig. \ref{chi-sgp=7-scale}.
All data for $\chi_{\mbox{\scriptsize SG}}$ with the exception of
$L=4$ scale well with $T_c(p=0.7)=0.85\:(0.05)$, $\nu=1.0\:(0.2)$ and
$\eta=0.1\:(0.2)$. The fact that the spin glass susceptibility
satisfies the above scaling form (\ref{scale-chi-sg}) strongly
suggests that the magnetic order of this model has changed between
$p=0.8$ and $p=0.7$ from antiferromagnet to spin glass.

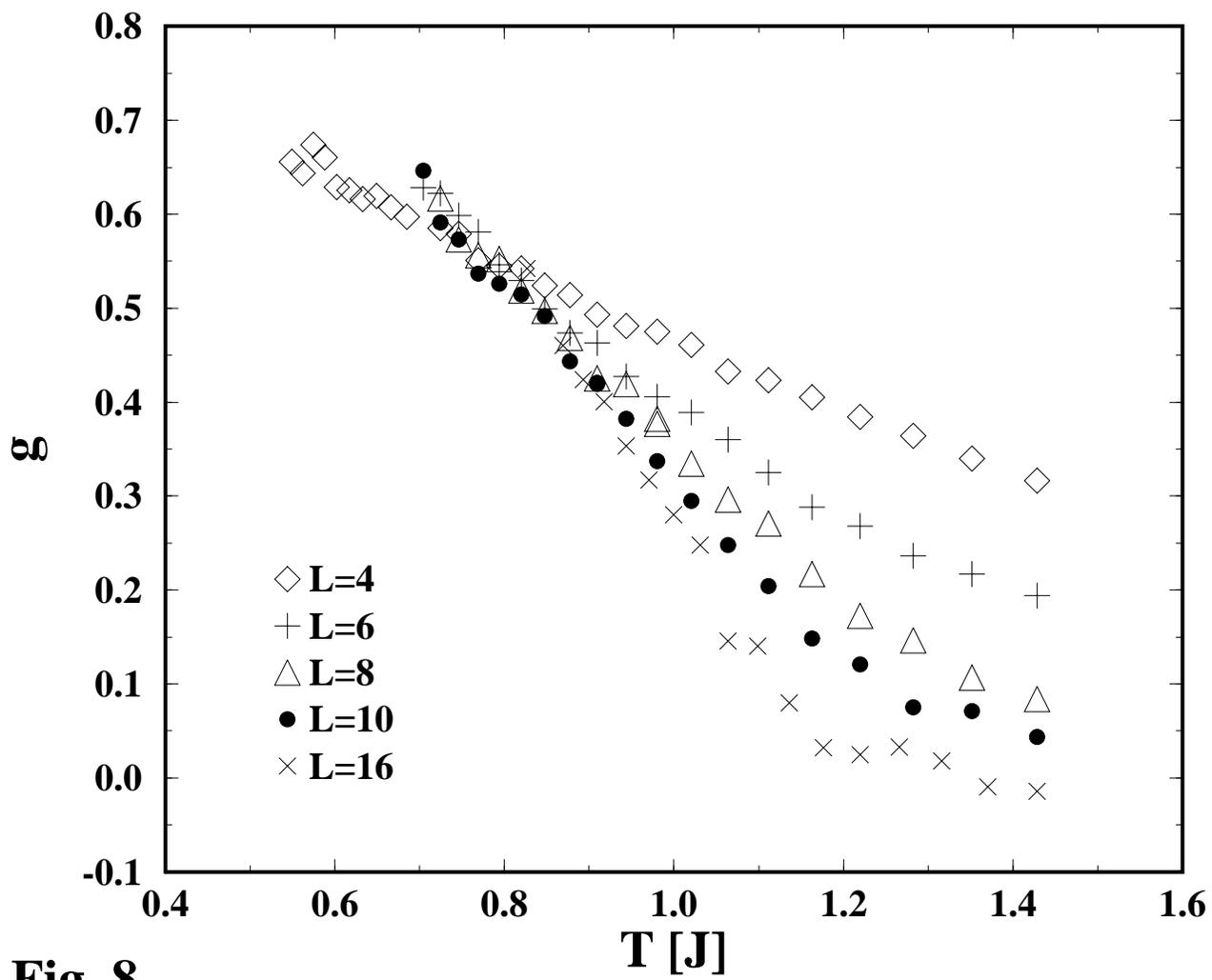
\begin{figure}
\caption{Binder cumulant of the spin glass order parameter versus
temperature, $p=0.7$.}
\label{binp=7}
\end{figure}

In Fig. \ref{binp=7} we present our data of $g$ at $p=0.7$.
The data show the typical behaviour as it has been observed in short
range spin glasses, i.~e., the data merge at approximately
$T=0.85$, indicating the phase transition. There is even a slight
tendency of fanning out of the data below this intersection point,
which is a strong evidence for the occurrence of a phase transition.
Such a fanning out is usually observed in uniform systems. In our
system this
may be due to the proximity to the tricritical point or to residual
short range antiferromagnetic order (see below).
The best fit was achieved
with $T_c(p=0.7)=0.83\:(0.05)$ and $\nu=1.05\:(0.2)$, which agrees well with
our estimates from the scaled spin glass susceptibility. The stronger
scattering of this quantity, especially of the
$\left[\langle q^4\rangle\right]$ data, may be due to insufficient
disorder averaging (40 systems for $L=16$) and could probably be
decreased with additional computational power.

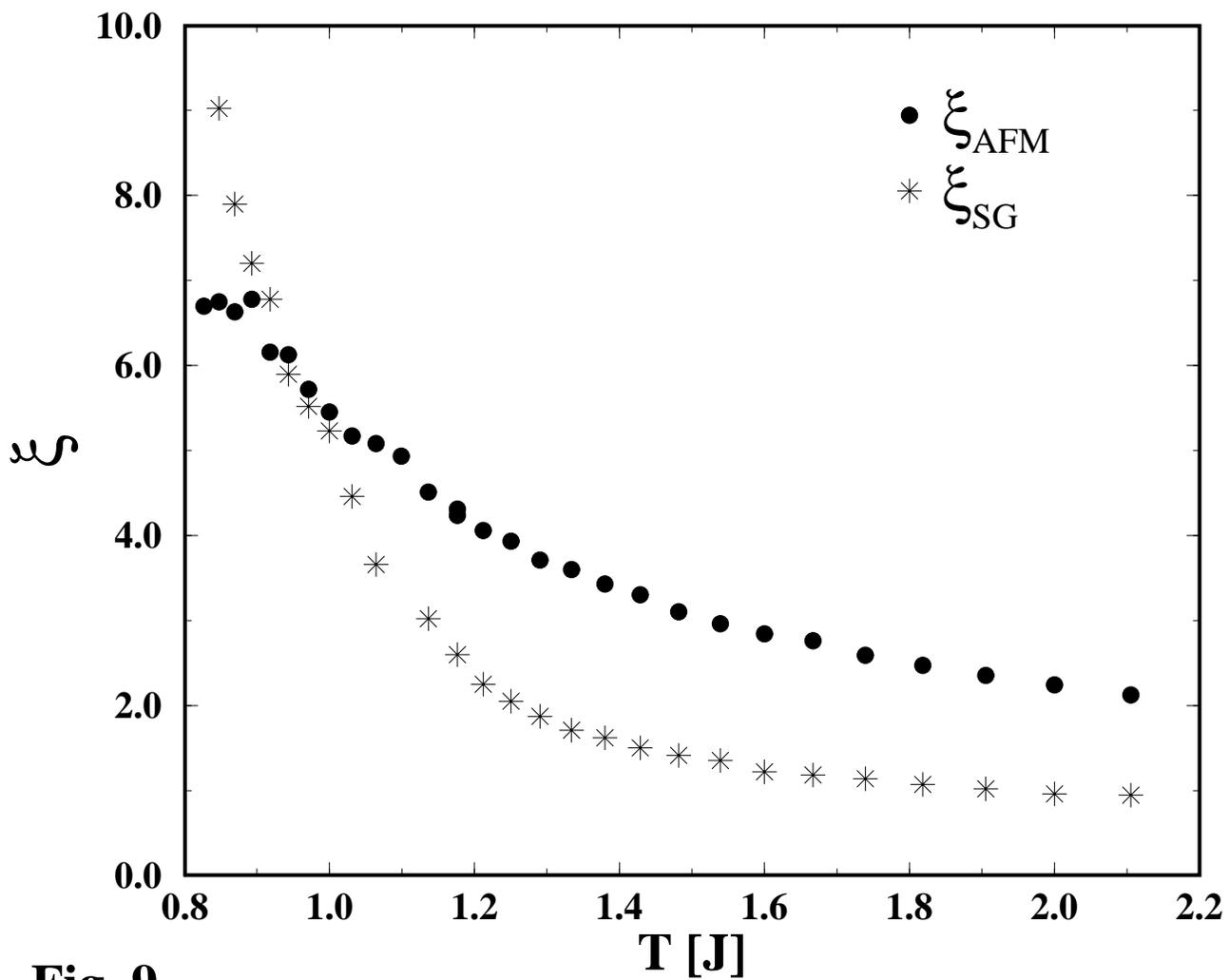
\begin{figure}
\caption{Spin glass correlation length $\xi_{SG}$
         ($\ast$) and antiferromagnetic correlation length
         $\xi_{AFM}$ ($\bullet$) versus
         temperature, $p=0.7$.}
\label{sg-cor-len}
\end{figure}

In Fig.~\ref{sg-cor-len}
the temperature dependence of $\xi_{\mbox{\scriptsize SG}}$
and $\xi_{\mbox{\scriptsize AFM}}$ are presented.
While the spin glass correlation length is very small for large $T$
but increases drastically as $T\to T_c\approx0.83$, the
antiferromagnetic correlation length starts at a higher level but
increases slower than $\xi_{\mbox{\scriptsize SG}}$; it ceases
to increase at about $T=0.9$ and saturates for lower temperatures.

All these results of our simulations at $p=0.7$ are consistent with the
interpretation that in this dilution regime we witness indeed a spin
glass transition, but also encounter antiferromagnetic order of long
but {\it finite} range being embedded into long range spin glass
order. Apparently, a change of magnetic order has taken place within
the interval $p_*\in(0.7,0.8)$, where $p_*$ denotes the critical
dilution, where this change happens.

\subsection{Results for strong dilution}

Additional simulations were performed for concentrations
$p=\{0.6,0.5,0.4,0.3\}$. With stronger dilution we find that the spin
glass phase already encountered for $p=0.7$ persists. Both, spin glass
susceptibility and Binder cumulant of the order parameter scale well
in this regime with slight dilution dependent critical
exponents. Also, in this regime
the data of the Binder cumulant stay together below $T_c$ for all sizes,
as has been observed in other short range Ising spin glass simulations.
Unfortunately, it is increasingly difficult in this regime to
equilibrate the systems. As the concentration is lowered, the critical
temperature decreases rapidly (as expected
from our argument earlier, that $T_c(p_c')=0$ with $p_c'>p_c=0.195$),
but the characteristic microscopic energy barriers remain of order
unity. Metastability becomes an increasingly important problem
and is prohibitive in large systems.
In our simulations, for $p<0.6$ we
were unable to equilibrate the $L=16$ systems sufficiently close to
$T_c$ within reasonable computer time.

{}From analysing the correlation function in the strong
dilution regime we find two results: First, the spin glass correlation length
shows a similar divergence close to the critical temperature for
$p=\{0.6,0.5,0.4\}$ as we have seen in the previous subsection,
confirming again the development of long range spin
glass order. Second, we still find short range antiferromagnetic
order, which decreases with lower concentration: for $p=0.6$
$\xi_{\mbox{\scriptsize AFM}}$ rises slowly when $T$ is
lowered and at $T_c$ we have $\xi_{\mbox{\scriptsize AFM}}\approx 4$;
for $p=0.4$ the correlation length remains constant at
$\xi_{\mbox{\scriptsize AFM}}\approx 2$ for all temperatures which we
can simulate ($T\in[0.56,1.0], T_c(p=0.4)\approx 0.47$).

Furthermore, both critical exponents $\nu$ and $\eta$ apparently
decrease with increasing dilution (see the fitted values in table~I).
The generic theoretical expectation is that the exponents should be universal
all along the spin glass transition line,
but the issue of universality is not completely settled in diluted
systems\cite{footnote:dilute-uni}.
In our case, we note that the drift in the exponents is pronounced
for $0.5<p<0.7$  but minimal for $0.5\ge p\ge0.3$.
Thus we suggest that this dependence is an artifact of the
antiferromagnetic correlation length,
which is large and $p$-dependent for $0.5<p<0.7$, but is
quite small for $0.3<p<0.5$.
However, we have insufficient information to check this proposition,
and shall come back to this issue in the conclusion.

We conclude, that for the whole range $p\in[0.3,0.7]$ our
simulations testify the existence of a spin glass phase transition
for the short range antiferromagnetic Ising model on a fcc lattice.
Simultaneously, the model also exhibits residual antiferromagnetic
order, that is relatively long ranged at $p=0.7$ and saturates close
to the critical temperature; it decreases with further dilution,
showing no temperature dependence for $p=0.4$.

\section{Conclusion}
\label{sec:conclusion}

In this paper we have presented a Monte Carlo simulation
of the diluted short range antiferromagnetic Ising model on a fcc
lattice. The main purpose was to investigate the thermodynamic
equilibrium properties of this model.

In the undiluted case ($p=1$) we have found a first order phase
transition to an antiferromagnetically ordered state, consistent with
earlier simulations. Upon slight dilution, the
antiferromagnetic order persists, the first order transition becoming
weaker and changing to a continuous transition at the tricritical
concentration $p_{\rm tri}\approx 0.85$. At $p=0.8$ we still find
antiferromagnetic order from analysis of the correlation function and
scaling of the staggered susceptibility. Together with scaling of the
heat capacity, we find mutually consistent critical exponents. The question of
universality remains unclear; in view of the proximity to
$p_{\rm tri}\approx 0.85$ it seems likely that we observe tricritical
exponents at this point.

Below $p=0.8$ the divergence of the spin glass correlation length
as well as scaling of the spin glass susceptibility and the Binder
cumulant of the order parameter signal spin glass order.
This means, that there must be a multicritical point at
some concentration $p_*\in(0.7,0.8)$, where the change of
magnetic order takes place.
Simultaneously, the quasi temperature independent staggered
magnetization and the saturation of the antiferromagnetic correlation
length suggest the breakdown of antiferromagnetic long range order.
The fact, that the antiferromagnetic correlation length saturates
while spin glass correlations still grow contradicts the view
of a dynamically inhibited transition to an
antiferromagnet. Rather, our data suggest that the coexistence of
antiferromagnetic short range order together with long range
established spin glass order seems to be a special phenomena of this
diluted model.

%%%\subsection{Universality?}

Let us now turn to the question of universality in the spin glass
phase. In this phase (see Sec.~\ref{sec:spin-glass-order})
our scaling fits yielded critical exponents slightly dependent on
dilution (compare table~\ref{exponents}); we speculated, that this
could be an artifact of the changing antiferromagnetic correlation
length. Even in
unfrustrated models, there is still controversy over the universality
of exponents under dilution\cite{footnote:dilute-uni}.
Although our simulations are less extensive than the
MC--simulations of the short range Ising spin glass by Bhatt
and Young and by Ogielski, it is interesting to compare our exponents
with their values for the $\pm J$--model in $d=3$.
For this model, they find\cite{bhatt85,ogielski85}
$\nu\approx1.2$ and $\eta\approx-0.25$.
These values are just at or slightly out of the error margins of the present
simulations, so that it is not entirely clear, whether the two models
lie in the same universality class.
Note also, that our value of $\gamma$ is smaller than for the
``classic'' spin glass simulations, roughly $\gamma\approx1.8$ for
$p\in[0.3,0.7]$.
Additional simulations of larger
systems would be desirable to confirm our result.

It should be noted that even the best current simulations
on hypercubic lattices have not put to rest the
basic question whether the lower critical dimension for
Ising spin glasses is below 3, or equal to
3\cite{bhatt85,ogielski85,bhatt88,hetzel93,marinetti94,bernardi94}.
At the more modest scale of our simulation,
it cannot be decisively answered whether the diluted fcc Ising
antiferromagnet has a genuine spin glass transition, and if so
whether it has precisely the same exponents as the standard example of
$\pm J$ spin glasses on hypercubic lattices using array processors.
However, our data is consistent with both of these propositions.

%%%\subsection{Comparison with experiment}

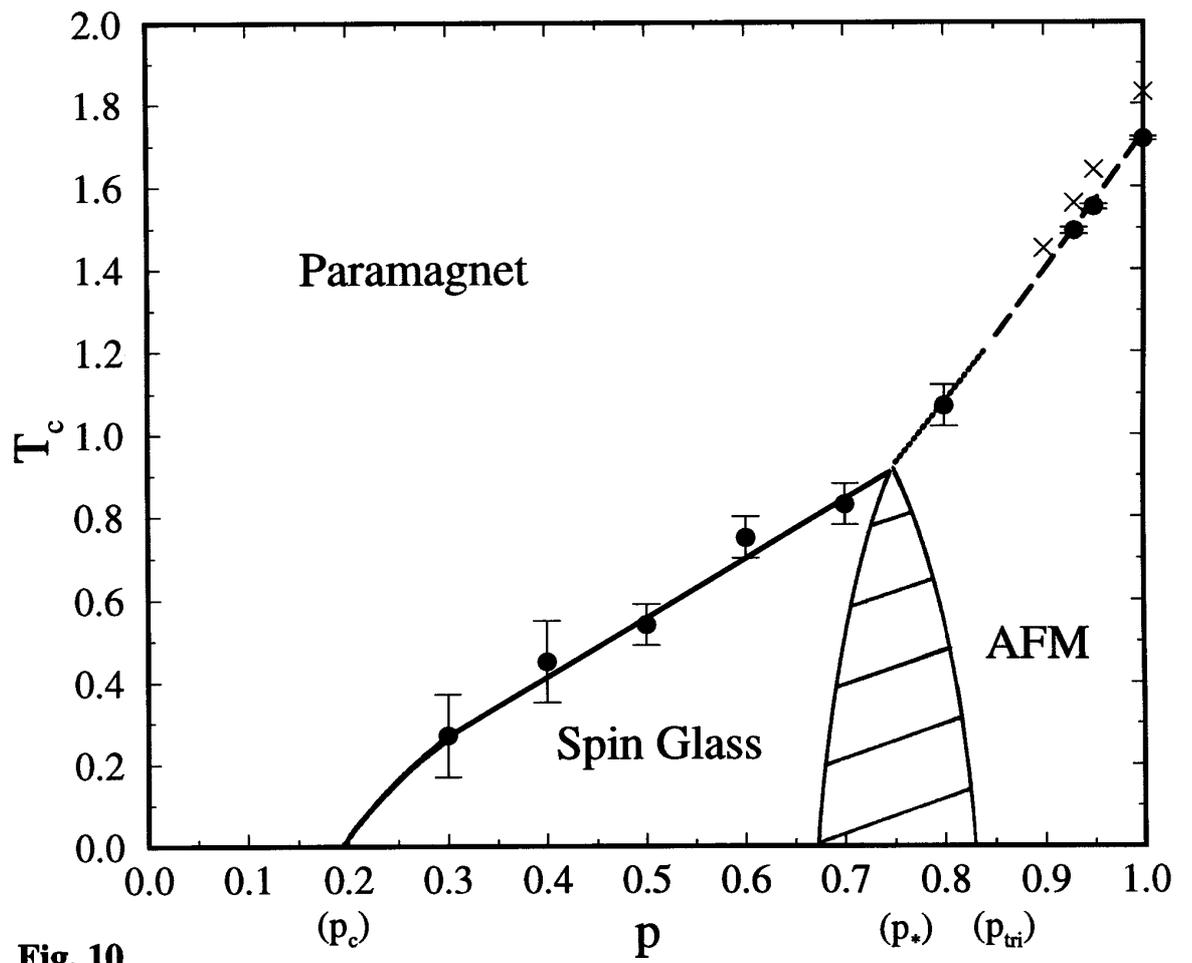
\begin{figure}
\caption{Phase diagram of the short range fcc Ising antiferromagnet
         with dilution. We plot $T_c$ versus spin concentration $p$.
         We observed a  $1^{st}$ order phase transition to an
         afm--state for $1.0\ge p\ge p_{tri} \approx 0.85$ (long
	 dashed line), becoming continuous for $p_{tri}\ge p\ge
         p_*\approx 0.75$ (dotted line). The dots above the transition
         line denote points ($T_c(L),p_{tri}(L)$)
         with the condition $Y=50$\% (see Sec.~IV).
         Upon further dilution
         ($p\le p_*$) we encountered a spin glass transition (full line,
         being extended down to the percolation threshold $p_c=0.195$).
         The lines serves only to guide the eye. The gothic arch
         marks a region, where the order of the system is unknown
         (afm, spin glass or coexistence of both).}
\label{phase-diag}
\end{figure}

Although in this work we have performed an Ising model simulation
we find striking similarities with experimental observations in DMS,
which have Heisenberg like magnetic moments\cite{heis-low-crit-dim}.
Not only are the respective phase diagrams (compare
Fig.~\ref{phase-diag} and Ref.~\onlinecite{giebultowicz93})
qualitatively similar and the various critical concentrations are
numerically close, but also the concentration dependence of
$\xi_{\mbox{\scriptsize AFM}}$ in the spin glass phase agrees well
with experiment. To be more specific, a comparison of our simulations
with intensive experimental studies by Giebultowicz
et~al.\cite{giebultowicz93} on ${\rm Zn_{1-p}Mn_pTe}$ shows the
following agreements:
(i) the afm transition is first order for $p\in[0.85,1.0]$, (ii) the
transition is continuous (to an afm long--range state) for
$p\in[0.75,0.85]$, (iii) the afm order is of large but finite
range for $p<p_*\approx 0.75$ in the spin glass phase and (iv)
the afm correlation length decreases with increasing dilution $p<p_*$.
These results together with the measured divergence of the nonlinear
susceptibility in ${\rm Cd_{1-p}Mn_pTe}$ (see
Ref.~\onlinecite{mauger89})
below $p_*\approx0.75$ do support
the view of a spin glass phase in experimental Heisenberg systems.

\begin{acknowledgments}
C.~W. and A.~Z. would like to thank Reiner Kree for
helpful discussions and J.~Holm for his assistance. C.~W. and A.~Z.
gratefully acknowledge support by SFB~345. C.~L.~H. is supported by
NSF grant DMR--9214943.
\end{acknowledgments}

%------------------------------------------------------------------
% Table 1

\renewcommand{\arraystretch}{0.8}
\begin{table}
\caption{Critical quantities for different concentration $p$ from
scaling of the respective susceptibility, Binder cumulant and
analysis of the respective correlation function. Not listed are the
scaling results of the heat capacity for $p=0.8$. They are:
$T_c=1.07\,(0.05)$, $\alpha=0.38\,(0.1)$ and $\nu=0.57\,(0.1)$. With
susceptibility and correlation function we denote the valid quantity
for the respective dilution regime.}
\label{exponents}
\begin{tabular}{cdddd}\hline
\multicolumn{5}{c}{\bf Critical Temperatures and Exponents}\\
\hline
Order & Concentration & Susceptibility & Binder cumulant &
Correlation function \\
\hline\hline
 & & $T_c$=1.07 (0.05) & &  \\
AFM & 0.8  &  $\nu$=0.51 (0.10) & & $\nu$=0.55 \\
 & & $\eta$=0.05 (0.15) & & $\eta=-$0.04 (0.05) \\ \hline\hline
 & & $T_c$=0.83 (0.05) & $T_c$=0.83 (0.05) & \\
SG & 0.7  &  $\nu$=1.00 (0.20) & $\nu$=1.05 (0.20) & $\nu$=0.96 \\
 & &  $\eta$=0.10 (0.20) &  &  $\eta$=0.15 (0.20)\\ \hline
 & & $T_c$=0.75 (0.05) & $T_c$=0.76 (0.05) & \\
SG & 0.6  &  $\nu$=0.80 (0.20) & $\nu$=0.90 (0.20) & $\nu$=0.98 \\
 & &  $\eta$=0.00 (0.20) & & $\eta$=0.07 (0.15)\\ \hline
 & & $T_c$=0.55 (0.05) & $T_c$=0.53 (0.05) & \\
SG & 0.5 & $\nu=$0.73 (0.20) & $\nu=$0.73 (0.20) & \\
 & & $\eta=-$0.30 (0.25) & & \\ \hline
 & & $T_c$=0.47 (0.10) & $T_c$=0.44 (0.10) & \\
SG & 0.4 & $\nu$=0.80 (0.20) & $\nu$=0.80 (0.20) & \\
 & & $\eta=-$0.35  (0.20) & & \\ \hline
 & &  $T_c=$0.27 (0.10) & $T_c=$0.27 (0.10) &  \\
SG & 0.3 & $\nu=$0.70 (0.20) & $\nu=$0.70 (0.25) &  \\
 & &  $\eta=-$0.40 (0.20) & & \\ \hline
\end{tabular}
\end{table}

%-----------------------------------------------------------------
%Here are all figure captions

\begin{thebibliography}{99}
\bibitem{twopstar}
  In principle, one could imagine a intermediate phase which had coexisting
  antiferromagnetic and spin glass order, which would thus have
  distinct critical concentrations $p_*$.
  That phase would be analogous to a state of two infinite interpenetrating
  percolation clusters, which is geometrically possible in $d=3$.
  Within the rather coarse $p$ grid of our simulations, however,
  only one $p_*$ value is observed along the line $T=T_c(p)$.
\bibitem{afmsgclass}
  It is intriguing to speculate on the universality class of
  the transition as a function of $p$ along the line $p_*(T)$.
  The afm order parameter would be expected to
  behave as in a random--field 3--state--Potts model (since the effective
  random fields distinguish among three ordering wavevectors)
  yet the spin glass order parameter would behave as at the
  spin-glass freezing temperature.
  Our simulations  cannot address this question
  (which is best studied at $T=0$).
\bibitem{errorbars}
  The listed errorbars of the critical exponents and the critical
  temperature are generous estimates from fitting the respective
  scaling form to our simulation data.
\bibitem{heis-low-crit-dim}
  It is commonly believed that $d=3$ is {\it below} the lower critical
  dimension for Heisenberg spin glasses, but anisotropic
  spin interactions --- which are certainly present in DMS ---  are
  believed to drive the transition to the Ising class, even though the
  ordered state may be held together mainly by the isotropic interactions.
\end{thebibliography}
\end{document}